\documentclass[aps,pra,superscriptaddress,twocolumn,longbibliography]{revtex4-2}

\usepackage{graphicx}
\usepackage{amsmath} 
\usepackage{amssymb}
\usepackage{hyperref}
\usepackage[utf8]{inputenc}
\usepackage{mathtools}
\usepackage[english]{babel}
\usepackage{bbm}
\hypersetup{colorlinks=true, linkcolor=blue, citecolor=blue, urlcolor=blue}
\usepackage{xcolor}
\usepackage{braket}
\usepackage{array}
\usepackage{soul}
\newcolumntype{P}[1]{>{\centering\arraybackslash}p{#1}}

\usepackage{dsfont}

\newcommand{\ie}{i.\,e.,\ }

\newcommand{\re}{\mathrm{Re}}
\newcommand{\im}{\mathrm{Im}}

\newcommand{\rr}{\mathbf{r}}

\newcommand{\fref}[1]{\text{Fig.}~\ref{#1}}

\newcommand{\eref}[1]{\text{Eq.}~\eqref{#1}}
\newcommand{\eeref}[1]{\text{Eqs.}~\eqref{#1}}

\begin{document}
\title{
Two-dimensional motion of an impurity under dynamic light-induced dipole forces in an atomic subwavelength array
}

\author{Samuel Buckley-Bonanno$^\ddagger$}
\email{sbuckleybonanno@g.harvard.edu}

\author{Stefan Ostermann$^\ddagger$}
\email{stefanostermann@g.harvard.edu}

\author{Yidan Wang}

\author{Susanne F. Yelin}

\affiliation{Department of Physics, Harvard University, Cambridge, Massachusetts 02138, USA}

\altaffiliation{These authors contributed equally to this work.}

\begin{abstract} 
Long-range dipole-dipole interactions in subwavelength arrays of quantum emitters involve virtual photon exchange processes that impart forces on the emitters due to the imposed photon recoil. We perform a semi-classical analysis of the dynamics of an impurity allowed to freely move through a subwavelength array of atoms in different parameter regimes. We numerically solve the coupled set of equations between motional and spin degrees of freedom to elucidate the possible impurity trajectories realizable in this system. We find that the impurity can maintain quasi-stable orbits within the plaquette for long times. The regions through which these orbits pass are strongly dependent on the chosen atomic transition dipole moment. We further provide intuition for our findings based on a simplified model, where the lattice dynamics is adiabatically eliminated. As a final point of analysis, we also take the motional degrees of freedom of the lattice atoms into account, and study the polaron-like excitation induced in the kinetic state of the lattice by the impurity.
\end{abstract}

\maketitle

\section{Introduction}

\begin{figure}[t]
    \centering
    \includegraphics[width = \columnwidth]{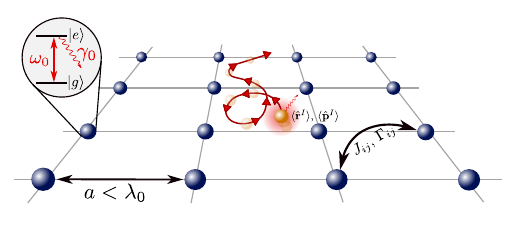}
    \caption{
    Sketch of the considered setup -- an impurity is placed at the center of a two-dimensional atomic array with subwavelength lattice spacing. Both the impurity and lattice atoms are assumed to be spin-1/2 particles with a transition frequency $\omega_0$ between the ground state $\ket{g}$ and excited state $\ket{e}$. The impurity interacts with the lattice and the lattice atoms with each other, all via light-induced coherent ($J_{ij}$) and dissipative ($\Gamma_{ij}$) couplings .
    }
    \label{fig:setup_sketch}
\end{figure}

\begin{figure}[t]
    \centering
    \includegraphics[width = \columnwidth]{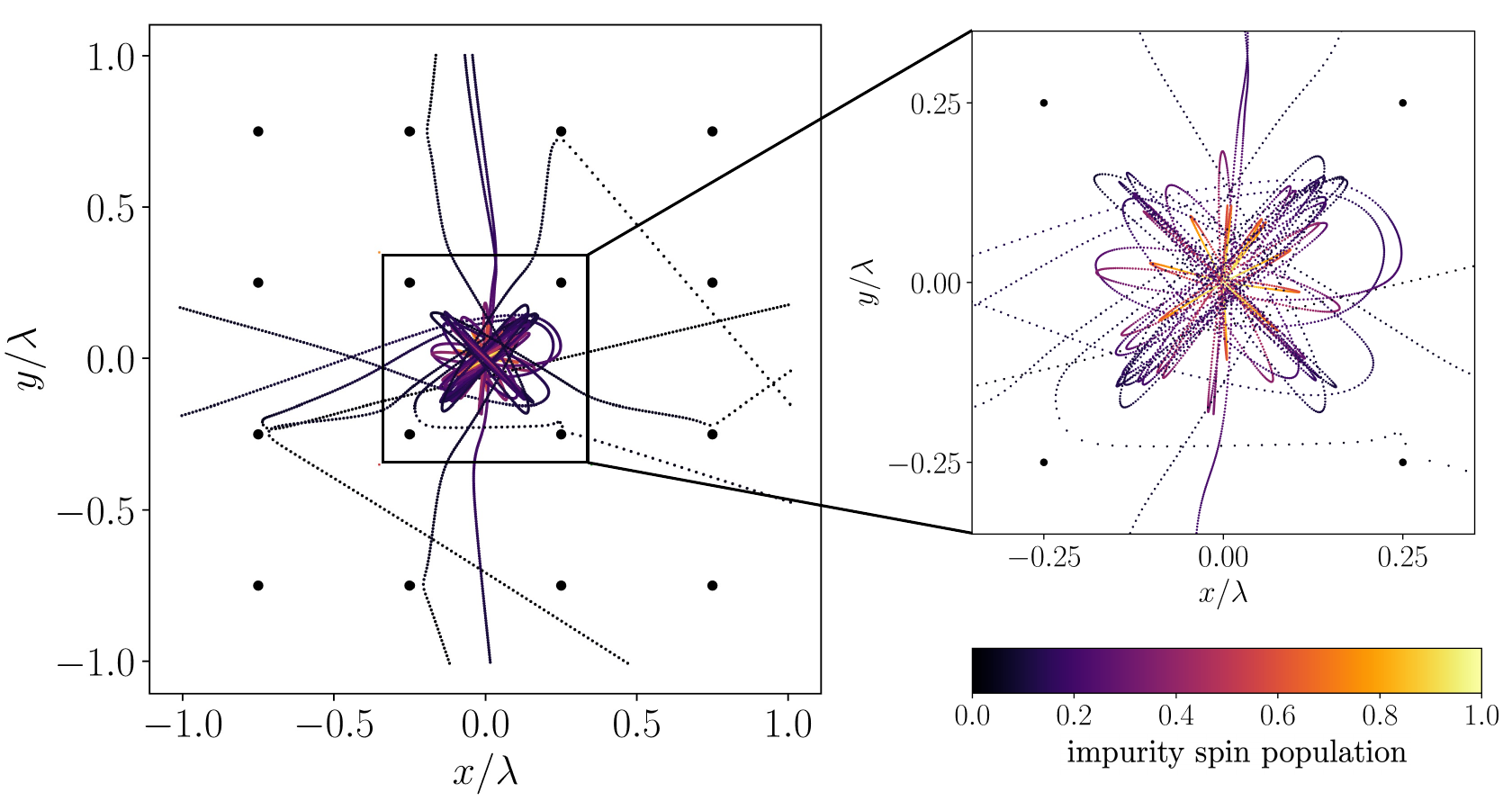}
    \caption{
Twelve exemplary orbits illustrating the variety of realizable trajectories for the impurity depending on the initialization. The impurity moves in the plane of a square lattice with lattice spacing $a=0.5\lambda_0$. 
For each trajectory the impurity is initialized in its excited state at the center of the lattice with an initial momentum of magnitude $|\mathbf{p}^I| = 0.05 \hbar k_0$. The initial direction of the impurity's motion is randomly chosen from a uniform distribution of angles. The color coding along each trajectory indicates the excited state population of the impurity. The distance between dots indicates the impurity's velocity. Small spacings between dots corresponds to slow impurity motion and large spacings to fast motion. The trajectories remain confined within the plaquette for multiple rounds and only escape the lattice once the excitation drops to low values. All trajectories are plotted on top of each other to give an example of typical impurity motion. 
Other parameters: $\omega_\mathrm{rec} = 2\pi^2\gamma_I$, $\delta_A = 0$, $\delta_I = 0$, $\gamma_I = \gamma_L$ and circular polarizations $\mathbf{d} = 1/\sqrt{2}(1,i,0)^T$ for the lattice atoms and the impurity.
    }
    \label{fig:trajectory_example}
\end{figure}
The tremendous progress in trapping and manipulating individual atoms in dense periodic geometries~\cite{lewenstein_ultracold_2007, bloch_quantum_2012, endres_atom-by-atom_2016, norcia_microscopic_2018, barredo_synthetic_2018, browaeys_many-body_2020} recently sparked interest in gaining a better understanding of cooperative phenomena like super- and subradiance in extended arrays of atoms~\cite{asenjo-garcia_exponential_2017, reitz_cooperative_2022}. The role of cooperativity in the dynamics of dense ensembles of atoms was first investigated decades ago by Dicke~\cite{dicke_coherence_1954,gross_superradiance_1982}. However, understanding the influence of the particular lattice geometry and the intricate effects tied to spatially inhomogeneous and distance-dependent interactions between quantum emitters in subwavelength geometries in free space~\cite{lehmberg_radiation_1970-1,lehmberg_radiation_1970-2}, is an ongoing research effort.

The motivation behind these efforts is two-fold. On the one hand, cooperative phenomena will play a role in any quantum dynamics studied in dense atom arrays, necessitating a deeper understanding and finer control of related effects~\cite{kramer_optimized_2016, hutson_observation_2024}. On the other hand, the availability of well controlled extended arrays of quantum emitters in free space enables an in depth systematic study of intriguing phenomena~\cite{rui_subradiant_2020, srakaew_subwavelength_2023}. The latter is particularly striking since cooperativity is at the core of many processes in nature on the nano-scale. Prominent examples are cooperative transport~\cite{needham_subradiance-protected_2019, masson_atomic-waveguide_2020, moreno-cardoner_efficient_2022, holzinger_harnessing_2023}, radiation storage based on subradiance~\cite{ballantine_subradiance-protected_2020,ferioli_storage_2021, rubies-bigorda_dynamic_2023} and superradiant emission enhancement~\cite{masson_many-body_2020, masson_universality_2022, sierra_dicke_2022, ferioli_non-equilibrium_2023,rubies-bigorda_characterizing_2023, yan_superradiant_2023}. In most of the cases studied so far, however, the atoms are assumed to be pinned down to their respective lattice positions, so that motional degrees of freedom are neglected. A recent series of works investigated the effect of motion along a single axis~\cite{palmer_enhancing_2010, chang_self-organization_2013, ostermann_scattering_2014, hotter_superradiant_2019, shahmoon_chapter_2019, shahmoon_quantum_2020, suresh_atom_2022, rubies-bigorda_collectively_2024} and pointed out that motion can play a significant role in the collective response of subwavelength arrays. However, a detailed study of motion along multiple axes is still outstanding.

In this work we study the motion of an impurity in a two-dimensional subwavelength lattice (see~\fref{fig:setup_sketch}). Impurities embedded in cooperative arrays, without taking motion into account, have attracted a lot of attention recently because collective lattice modes can be harnessed modify the impurities' radiative properties and coupling to the free space electromagnetic vacuum~\cite{patti_controlling_2021, buckley-bonanno_optimized_2022, moreno-cardoner_efficient_2022}. Here, we study a semi-classical model where the impurity can move freely in the lattice plane. We show that light-induced dipole forces emerging due to virtual photon exchange processes between the lattice atoms and the impurity result in intriguing impurity dynamics depending on the initial conditions.
The interaction of the impurity with the lattice atoms results in a dynamically generated effective potential for the impurity. We determine stationary impurity positions within the lattice plane and investigate the role of the initial conditions for a broad range of configurations. Adiabatically eliminating the lattice gives rise to a closed-form expression for the effective potential which provides intuitive insights into the dynamics. Finally, we also investigate the role of motional degrees of freedom for the lattice atoms within their respective harmonic trapping potential. We show that this alters the impurity motion, which gets ``dressed'' by the lattice motion, reminiscent of polaron dynamics in solid state systems.

\section{Model}
The Hamiltonian describing the dynamics of the system can be decomposed into different parts via
\begin{equation}
    H = H_L + H_I + H_\mathrm{int}^L + H_\mathrm{int}^{LI} + H_\mathrm{kin}^I,
    \label{eqn:Hamiltonian}
\end{equation}
where $\hat{H}_L = (\omega_L-i\gamma_L/2) \sum_{i=1}^{N_L} \hat{\sigma}_i^\dagger \hat{\sigma}_i$ is the Hamiltonian describing the $N_L$ individual two-level systems making up the lattice and $\hat{H}_I = (\omega_I - i\gamma_I/2) \hat{s}^\dagger \hat{s}$ is the equivalent for the single impurity. We introduced the transition frequency $\omega_L \equiv \omega_e^L - \omega_g^L$ ($\omega_I \equiv \omega_e^I - \omega_g^I$) for the lattice (impurity) and the transition operators $\hat{\sigma}_i = \ket{g_i}\bra{e_i}$ ($\hat{s} = \ket{e_I}\bra{g_I}$) which describe transitions between the excited and ground states of the lattice atoms (the impurity). The finite lifetime of the excited state for the lattice $\gamma_L$ and the impurity $\gamma_I$ is included via the non-Hermitian part in the Hamiltonians. We choose $\hbar = 1$ throughout this work.

The electromagnetic vacuum field mediates coherent and dissipative long-range interactions between the lattice atoms at positions $\mathbf{r}_i$ via
\begin{subequations}
\begin{align}   
    H_\mathrm{int}^L &= \sum_{i,j \neq i}^{N_L} \left( J_{ij}(\mathbf{r}_i,\mathbf{r}_j) - \frac{i}{2} \Gamma_{ij}(\mathbf{r}_i,\mathbf{r}_j) \right) \sigma_i^\dag \sigma_j,\\
    &\equiv \sum_{i,j\neq i}^{N_L} \mathcal{C}_{ij}(\mathbf{r}_i,\mathbf{r}_j)\sigma_i^\dag \sigma_j,
\end{align}
\label{eqn:interaction_Hamiltonian_lattice}%
\end{subequations}
where $J_{ij}(\mathbf{r}_i,\mathbf{r}_j)$ ($\Gamma_{ij}(\mathbf{r}_i,\mathbf{r}_j)$) denotes the coherent (dissipative) pairwise interaction strength between lattice atoms at positions $\mathbf{r}_i$ and $\mathbf{r}_j$. We introduce the complex coupling constants $\mathcal{C}_{ij} \in \mathbb{C}$ for brevity. Similarly the interaction between the lattice atoms and the impurity at position $\mathbf{r}^I$ can be written as
\begin{equation}
        H_\mathrm{int}^{LI} = \sum_{i=1}^{N_L} \Big[\mathcal{C}_{iI}(\mathbf{r}_i,\mathbf{r}^I) \sigma_i^\dag s+ \mathcal{C}_{Ii}(\mathbf{r}^I,\mathbf{r}_i)s^\dag \sigma_i \Big],
\label{eqn:interaction_Hamiltonian_impurity}%
\end{equation}
where $\mathcal{C}_{iI}$ denotes the interaction between the lattice atoms and the impurity. The position dependence of the coherent and dissipative couplings is determined via
\begin{subequations}
    \begin{align}
    J_{mn}(\rr_m,\rr_n) &= -\frac{3\pi \sqrt{\gamma_m \gamma_n}}{\omega} {\mathbf{d}_m^\dagger} \cdot \re [\textbf{G}(\textbf{r}_{mn}, \omega)] \cdot \mathbf{d}_n,
    \label{eqn:J} \\
    \Gamma_{mn}(\rr_m,\rr_n) &= \frac{6\pi \sqrt{\gamma_m \gamma_n}}{\omega} {\mathbf{d}_m^\dagger} \cdot \im [\textbf{G}(\textbf{r}_{mn},\omega)] \cdot \mathbf{d}_n,
    \label{eqn:Gamma} 
    \end{align}
    \label{eqn:couplings}%
\end{subequations}
where $\gamma_m$ and $\gamma_n$ 
are the individual atoms' excited state decay rates, $\omega = c k_0 = 2\pi c/\lambda_0$ is the atomic transition frequency, where $\lambda_0$ is the wavelength associated with the transition frequency $\omega$.  $\mathbf{d}_{m,n}$ are the dipole moments of atoms $m$ and $n$, and $\mathbf{r}_{mn} = \mathbf{r}_m - \mathbf{r}_n$ is the vector connecting the two atoms. $\mathbf{G}(\mathbf{r}_{mn},\omega)$ denotes the Green's tensor for a point dipole in free space (see Appendix~\ref{app:Greens_Tensor}). Note that we use $\omega_L \approx \omega_I \equiv \omega$ to determine the coupling constants in~\eref{eqn:couplings} throughout this work. The Green's tensor formalism holds if $|\omega_I - \omega_L| \ll \omega_I , \omega_L$~\cite{gruner_green-function_1996, dung_resonant_2002}. The non-Hermitian Hamiltonians defined in~\eeref{eqn:interaction_Hamiltonian_lattice} and \eqref{eqn:interaction_Hamiltonian_impurity} only describe the system's dynamics in the single excitation subspace, where only a single atom is excited at a time, and the analysis presented below is restricted to this case. Embedding the impurity in a dense array of atoms can result in an extension of the effective excited state lifetime by orders of magnitude~\cite{patti_controlling_2021, buckley-bonanno_optimized_2022}. This feature is crucial for maintaining the impurity in an excited state for extended times, allowing the emergence of the complex dynamics described below.

The kinetic degrees of freedom for the impurity are included via the Hamiltonian $H_\mathrm{kin}^I = (\hat{\mathrm{p}}^I)^2/(2m)$. We move into a frame rotating at the lattice and impurity frequency, and define $ \tilde{\sigma}_i = \hat{\sigma}_i e^{i\omega_L t} $ and $\tilde{s} = \hat{s} e^{i \omega_I t}$ respectively. The Heisenberg equations of motion for the transition operators $\tilde{\sigma}_i$ and $\tilde{s}$ are then given as
\begin{align}
\partial_t{\tilde{s}} &= \left( i \delta_I + \frac{\gamma_I}{2} \right) \tilde{s} + i \sum_{i = 1}^{N_L} \left[\mathcal{C}_{Ii}(\mathbf{r}^I,\mathbf{r}_i) \tilde{\sigma}_i
+ \mathcal{C}_{iI}(\mathbf{r}_i,\mathbf{r}^I) \tilde{\sigma}_i^\dagger \right], \label{eqn:s_dynamics}\\
    \partial_t{\tilde{\sigma}}_i &= \left( i \delta_A + \frac{\gamma_A}{2} \right) \tilde{\sigma}_i + i \left[\mathcal{C}_{iI}(\mathbf{r}_i,\mathbf{r}^I) \tilde{s} + \mathcal{C}_{Ii}(\mathbf{r}^I,\mathbf{r}_i) \tilde{s}^\dag \right] \nonumber \\ 
    &\quad \quad \quad \quad \quad \quad \quad \, \, \, \, \, + i\sum_{j \neq i}^{N_L} \mathcal{C}_{ij}(\mathbf{r}_i,\mathbf{r}_j) \tilde{\sigma}_j.\label{eqn:sigma_dyn}
\end{align}
The equations of motion for the components of the impurity's position and momentum operators take the form,
\begin{subequations}
\begin{align}
\partial_t \hat{r}_\tau^I &= \hat{p}_\tau^I/m,\\
\partial_t{\hat{p}}_\tau^I &= \sum_{i=1}^{N_L} \Big\{ \tilde{s}^\dagger \left[\partial_\tau \mathcal{C}_{Ii}(\mathbf{r}^I,\mathbf{r}_i) \right] \tilde{\sigma}_i + \mathrm{h.c.}\Big\},\label{eqn:p_dyn}
\end{align}
\label{eqn:x_p_dyn}
\end{subequations}
for $\tau \in \{x,y,z\}$~\cite{shahmoon_chapter_2019, shahmoon_quantum_2020}. For the analysis below we use the recoil momentum $\hbar k_0$ with $k_0 = k_L$ as the fundamental unit for the momentum and the recoil energy and it's associated frequency as $E_\mathrm{rec} = \hbar \omega_\mathrm{rec} = \hbar^2 k_0^2/(2m)$ as the fundamental energy scale.

The analysis presented below is performed in the semi-classical limit, which means that the impurity is treated as a classical point particle, while the spin dynamics are treated quantum mechanically and governed by~\eeref{eqn:s_dynamics} and~\eqref{eqn:sigma_dyn}. This limit is obtained from the equations of motion presented above by replacing the position and momentum operators with their expectation values,~\ie $\hat{\mathbf{r}}^I\rightarrow\langle\hat{\mathbf{r}}^I\rangle \equiv \mathbf{r}^I$ and $\hat{\mathbf{p}}^I\rightarrow\langle\hat{\mathbf{p}}^I\rangle \equiv \mathbf{p}^I$. Note that the dynamics of these expectation values depend on the correlators $\langle\tilde{s}^\dagger \tilde{\sigma}_i\rangle$ and $\langle\tilde{\sigma}_i^\dagger \tilde{s}\rangle$ for $i=1...N_L$. Since we only consider impurity motion within the lattice plane and assume that the lattice spans the $x$-$y$ plane, we choose $p^I_z = r^I_z = 0$. This confinement can, for example, be achieved by adding an additional strong optical lattice along the $z$-direction, and then aligning one of its potential minima with the lattice plane. In this case, the impurity motion is mostly confined to this plane.

\begin{figure}
    \centering
    \includegraphics[width = 0.9\columnwidth]{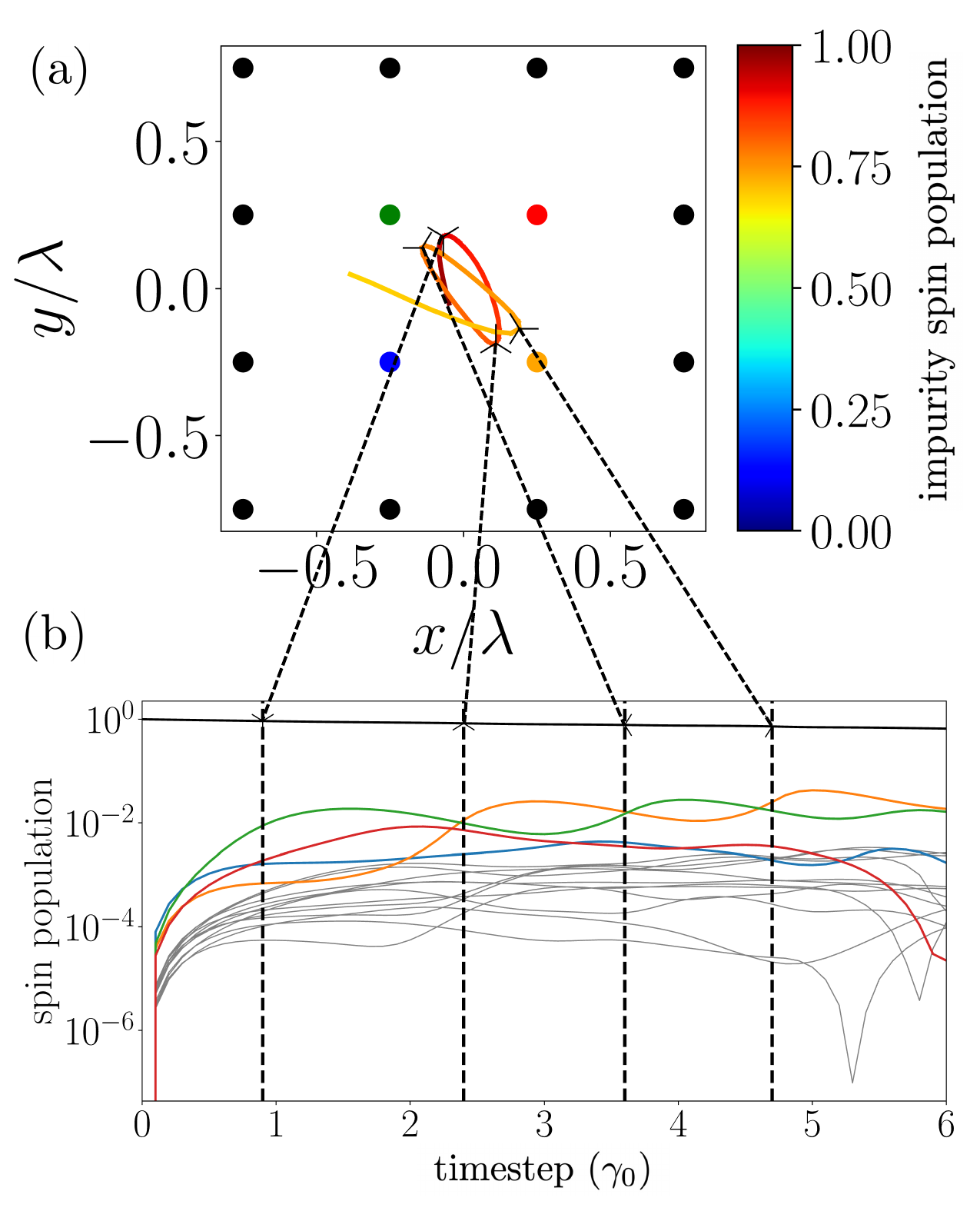}
    \caption{
    Connection between dynamics of external (motional) and internal (spin) degrees of freedom for a single trajectory. (a) The early time dynamics of a single trajectory of an impurity moving in a $4 \times 4$ lattice with lattice spacing $a=0.5\lambda_0$. The color coding along the trajectory of the impurity indicates its excited state population over time, as described by the color scale on the right. The colors on the fixed lattice points, on the other hand, indicate which lattice atoms are associated with the excited state population curves on the lower part of the figure. The initial position of the impurity is at $r^I_0 = (-0.1,-0.1,0)^T a$, and the initial momentum is $\mathbf{p}^I_0 = 0.05 [\cos(\theta),\sin(\theta),0]^T\hbar k_0$ with $\theta = 0.58\pi$. All other parameters are the same as in~\fref{fig:trajectory_example}. The impurity is initially on a stable orbit within the central plaquette. However, dissipation results in wider and wider circles as population leaves the impurity and its coupling with the surrounding lattice diminishes. Eventually, the impurity exits the lattice. (b) The excited state populations of the impurity (black line on the top), and the four lattice points constituting the central plaquette, as a function of time. The color of the spin population curve corresponds to the colors of the lattice dots shown in panel (a). The grey lines indicate the excited state populations of the other lattice atoms shown as black dots in panel (a). The dotted vertical lines highlight the excited state populations at the time corresponding to the marked positions on the impurity trajectory in panel (a). Transfer of population follows an oscillatory pattern, due to the quasi-periodic trajectory of the impurity itself.
    }
    \label{fig:single_traj_example}
\end{figure}

\section{Impurity trajectories}
\begin{figure*}[t]
\centering
\includegraphics[width=1.00\textwidth]{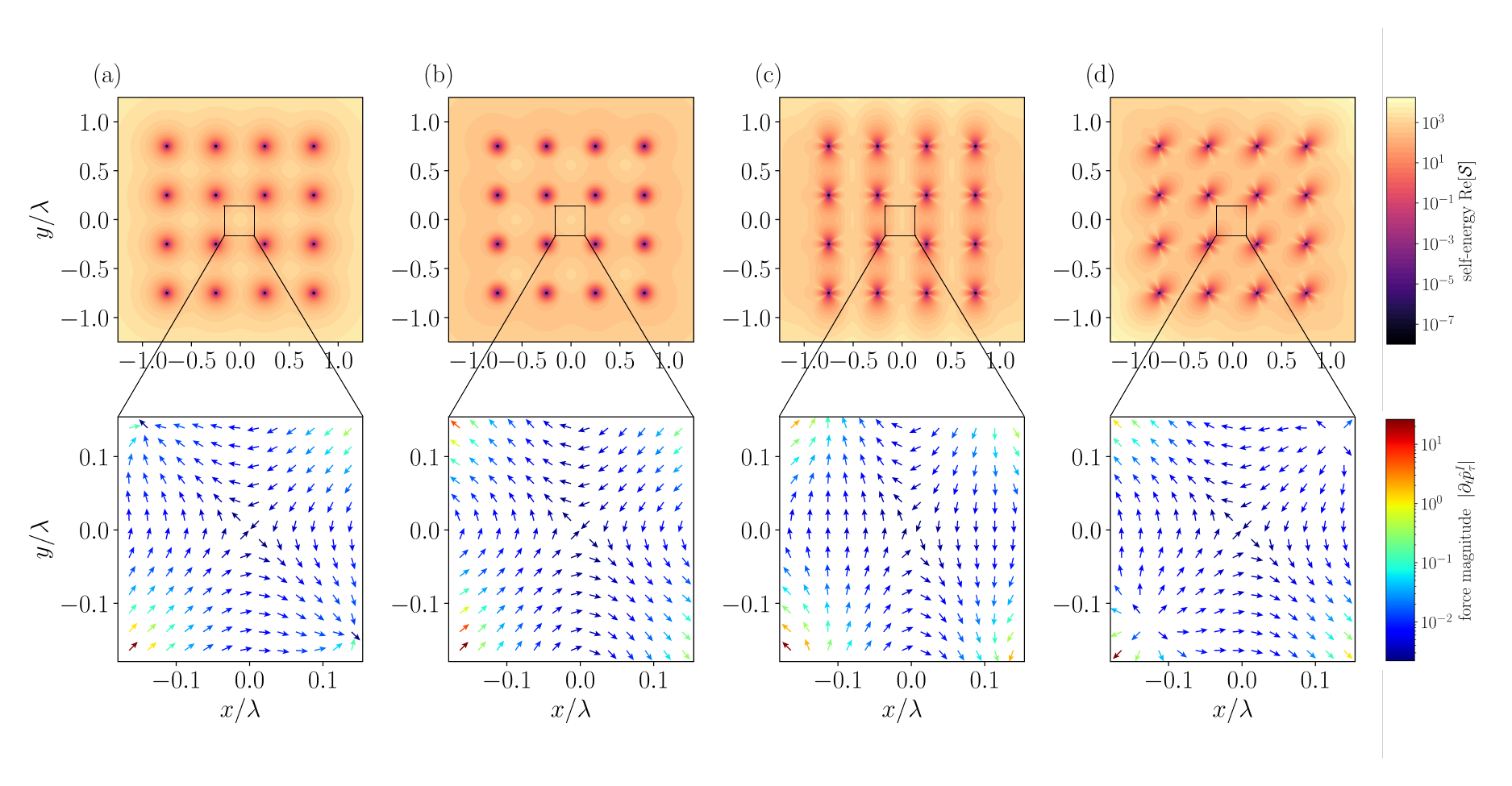}
\caption{
Illustration of the self-energy $\mathcal{S}$ defined below~\eref{eqn:imp_dyn_elim_lattice} (upper row) and the structure of the force vector $\partial_t \hat{p}_\tau^I$ calculated via~\eref{eqn:x_p_dyn_elimlatt} for different polarization directions $\mathbf{d}$ (lower row).
Note that the magnitude of the self-energy directly affects the strength of the force imparted to the impurity by the lattice atoms. (a) circular polarization $\mathbf{d} = 1/\sqrt{2}(1,i,0)^T$, (b) linear polarization perpendicular to the plane along the $z$-axis $\mathbf{d} = (0,0,1)^T$ (c) linear polarization along the $x$-axis $\mathbf{d} = (1,0,0)^T$,   and (d) linear polarization along the diagonal $\mathbf{d} = 1/\sqrt{2}(1,1,0)^T$. For spherical and perpendicular polarization the energy shift experienced by the impurity is rotationally symmetric around each lattice point, whereas for linear polarizations the structure is more complicated. All other parameters are the same as in previous figures.
}
\label{fig:effective_potential}
\end{figure*}
\eeref{eqn:s_dynamics} - \eqref{eqn:x_p_dyn} showcase the coupled non-linear dynamics between spin and motional degrees of freedom even in the semi-classical case.
The photon exchange between the impurity and the lattice atoms imposes forces $\propto \left[\partial_\sigma \mathcal{C}_{Ii}(\mathbf{r}^I,\mathbf{r}_i) \right] \langle\tilde{s}^\dagger \tilde{\sigma}_i \rangle + \text{h.c.}$
onto the impurity. To understand the impact of this on the impurity's dynamics, we numerically solve the semi-classical equations of motion self-consistently (see Appendix~\ref{app:num_solution}). Therefore, we initialize the impurity in its excited state at different positions within the central lattice plaquette of a square lattice with a given initial velocity. To illustrate the diversity in obtainable trajectories we show 12 different trajectories in~\fref{fig:trajectory_example} plotted on top of each other for an initial velocity in different directions in the $x$-$y$ plane. For this particular parameter choice the light-induced dipole forces are strong enough to confine the impurity motion to the central lattice plaquette, where it circles the center for several orbits. The radius of these orbits gets larger over time since dissipation slowly diminishes the impurity excitation, resulting in weaker forces between the lattice atoms and the impurity. This ultimately results in loss of the impurity, as evinced by the almost straight lines exiting the array in~\fref{fig:trajectory_example}.

To better understand the connection between the impurity motion and the spin dynamics, we show a single trajectory together with the excited state population dynamics of the impurity and the lattice atoms in~\fref{fig:single_traj_example}. As discussed above, the impurity exhibits approximately circular motion. The spin poplulations show that whenever the impurity approaches a lattice point, the spin population transfer between the impurity and the lattice point intensifies. As one can see from~\eref{eqn:p_dyn}, the force felt between the impurity and any particular lattice point is dependent on the expectation values $\langle\tilde{s}^\dagger \tilde{\sigma}_i \rangle$ and $\langle\tilde{\sigma}_i^\dagger \tilde{s} \rangle$. That implies that a force can only emerge if both, the lattice and impurity atom, have a finite excited state population. This implies that the initial momentum must be small enough for a sufficiently large transfer of spin population to occur between them, before the impurity passes away from the lattice point. This small momentum gives the lattice atom enough time to exert a force on the impurity, redirecting it back towards the center of the plaquette. Note that the fluctuations in excited state population shown in~\fref{fig:single_traj_example}(b) are at the frequency of the oscillatory motion of the impurity, with a phase determined by the position of the impurity within the plaquette relative to all four of its lattice points. The fluctuations in excited state populations are higher for the impurities that are more frequently approached by the impurity (indicated as green and yellow points in~\fref{fig:single_traj_example}. In this way, spin population transfer induced by long-range dipole-dipole interactions keeps the particle to perform orbits within the central plaquette, guided by an effective emerging potential. Note however, that no external control is required to produce a restoring force for drawing the impurity back to the center of the plaquette. 

At any given moment, the electromagnetic potential landscape for an impurity placed at the center of the plaquette is necessarily unstable, as required by Earnshaw's theorem from classical electrodynamics. However, the non-trivial interplay between impurity motion and spin dynamics ensures that this static potential is dynamically adjusted to result in multiple quasi-periodic orbits as shown in Figs.~\fref{fig:trajectory_example} and~\fref{fig:single_traj_example}. To some extent, this behavior resembles the micromotion of charged particles moving in a Paul trap where the potential is generated via fast-varying RF fields. However, unlike in these dynamical traps, where the field spins on a timescale much faster than the ion's motion, the frequencies of the excited state population curves shown in~\fref{fig:single_traj_example}(b) are in resonance with the motion of the impurity.

\section{Dynamic stability}
The dynamic stability and the shape of the force field experienced by the impurity within a lattice plaquette can be further analyzed by adiabatically eliminating the lattice degrees of freedom. This only captures the dynamics correctly if the spin dynamics of the lattice is much faster than the impurities motion. Nevertheless, this allows us to derive an effective force field that provides some intuition for the impurity dynamics in the general cases presented above.

Let $z = i \delta_I + \frac{\gamma_I}{2}$ and $ Z_{ij} = i J_{ij} + \frac{1}{2} \Gamma_{ij} $. The equation of motion for $\sigma_i$ can then be written as $\partial_t{\sigma}_i = z \sigma_i + Z_{is} s + \sum_{j \neq i} Z_{ij} \sigma_j$. Let $\sigma_i(t) = \sigma_i(\omega) e^{i\omega t}$, which results in $0 = (z - i\omega) \sigma_i + Z_{is} s + \sum_{j \neq i} Z_{ij} \sigma_j$.
This can be written in matrix form with $ \vec{\sigma} = \begin{pmatrix}
    \sigma_1 &
    \sigma_2 &
    \cdots &
    \sigma_N
\end{pmatrix}^T $ and $ \vec{Z}_s = \begin{pmatrix}
    Z_{s1} & Z_{s2} & \cdots & Z_{sN}
\end{pmatrix}^T $, as well as
\begin{equation}
\tilde{Z}_0 = \begin{pmatrix}
    0 & Z_{21} & Z_{31} & \cdots &  Z_{N1} \\ 
    Z_{12} & 0 & Z_{32} & & \vdots \\
    Z_{13} & Z_{23} & 0 \\ 
    \vdots & & & \ddots & \\ 
    Z_{1N} & \cdots & &  & 0
\end{pmatrix},
\end{equation}
so that the equations of motion~\eref{eqn:sigma_dyn} may be formulated as 
\begin{equation}
\vec{\sigma} = - \left( (z - i\omega) \mathds{1}  + \tilde{Z}_0 \right)^{-1} \vec{Z}_s s.
\end{equation}
Plugging this into~\eref{eqn:s_dynamics} and  using $s(t) = s(\omega) e^{i\omega t}$ results in
\begin{equation}
i \omega s = \left( z - \vec{Z_s}^\dag \left( (z - i\omega) \mathds{1}  + \tilde{Z}_0 \right)^{-1} \vec{Z}_s \right) s.
\label{eqn:imp_dyn_elim_lattice}
\end{equation}
From this equation we can read off the effect of the lattice onto the impurity via the apparent self-energy $\mathcal{S}\equiv -\vec{Z_s}^\dag \left( (z - i\omega) \mathds{1}  + \tilde{Z}_0 \right)^{-1} \vec{Z}_s = \vec{Z_s}^\dag\cdot \mathcal{M} \cdot \vec{Z_s}$. Plugging this into~\eref{eqn:x_p_dyn} we get
\begin{equation}
\partial_t{\hat{p}}^I_\tau = -i \Big\{ - s^\dagger \left[\left(\partial_\tau \vec{Z}_{s} \right) \cdot  \mathcal{M} \cdot \vec{Z}_s \right]s
+ \text{h.c.} \Big\}.
\label{eqn:x_p_dyn_elimlatt}
\end{equation}
In the semi-classical limit where the impurity is assumed to be a classical point particle the spatial structure of the force experienced by the impurity at a certain position within the lattice is determined by the term $\propto \left[\left(\partial_\tau \vec{Z}_{s} \right) \cdot  \mathcal{M} \cdot \vec{Z}_s \right]$, whereas the ultimate magnitude of the force depends on the impurity's excited state population $\langle s^\dag s\rangle$.

\begin{figure}
\centering
\includegraphics[width=0.96\columnwidth]{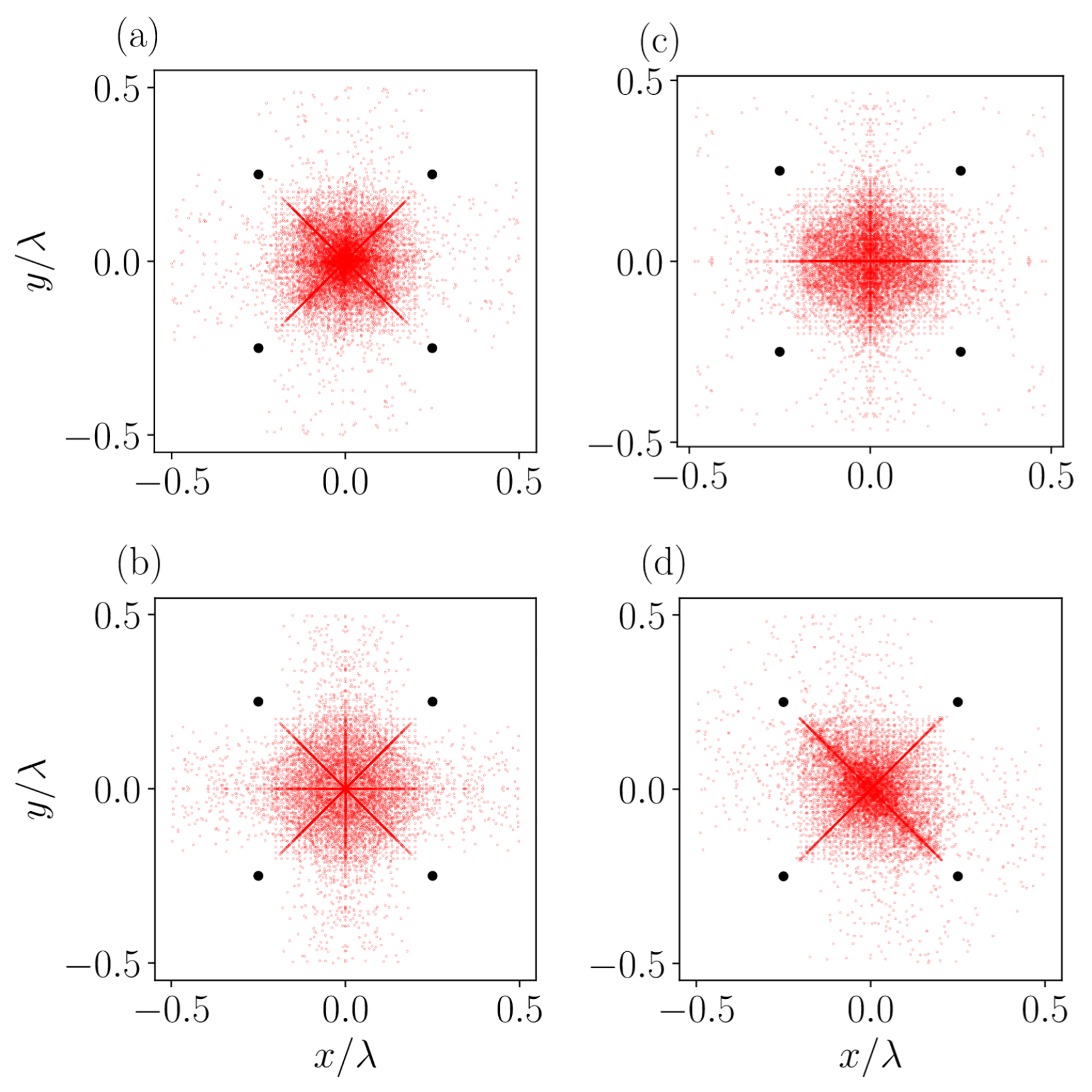}
\caption{Impurity position distributions for trajectories initialized across the central plaquette with zero initial momentum. Particles are initially placed with $\hat{\mathbf{p}}^I = 0$ in $25 \times 25$ different positions, uniformly distributed inside a square that reaches from $0.1 a$ to $0.9 a$ within the spacing of the lattice. They are then allowed to move throughout the plaquette, and points from their trajectories are overlaid. This provides a density plot indicating the mostly likely impurity positions while it moves throughout the plaquette. (a) circular polarization $\mathbf{d} = 1/\sqrt{2}(1,i,0)^T$, (b) linear polarization perpendicular to the lattice plane $\mathbf{d} = (0,0,1)^T$, (c) linear polarization along the $x$-axis $\mathbf{d} = (1,0,0)^T$, and (d) in-plane linear polarization along the diagonal $\mathbf{d}=1/\sqrt{2}(1,1,0)^T$. In addition to the stable point at the center, diagonal lines between lattice points also contain stable oscillatory trajectories. When impurities exit the lattice, it is typically along narrow lines that are aligned with the lattice structure, along axes that pass through the center. All other parameters are the same as in the previous figures.
}
\label{fig:full_stability}
\end{figure}
\subsection{Effective dynamics}
In~\fref{fig:effective_potential}, the upper row shows the position dependence of the self-energy $\mathcal{S}$. 
This complex quantity imposes an additional energy shift and dissipation term in~\eref{eqn:imp_dyn_elim_lattice}. This shift will result in a change of the impurity's excited state population when the impurity is located at different positions within the lattice and hence will influence the force felt by the impurity as described by~\eref{eqn:x_p_dyn_elimlatt}. These effective forces that also take the spatial structure $\propto \left[\left(\partial_\tau \vec{Z}_{s} \right) \cdot  \mathcal{M} \cdot \vec{Z}_s \right]$ into account are plotted in the lower row of~\fref{fig:effective_potential}. We find that different choices of polarization modify both the spatial structure of these energy shifts and the resultant force fields. For circular polarization~\fref{fig:effective_potential}(a) and for out-of-plane polarization~\fref{fig:effective_potential}(b), it exhibits a rotationally symmetric structure around each lattice point, whereas for in-plane linear polarization at different angles, the shift picks up a more complicated spatial structure [see~\fref{fig:effective_potential}(c)-(d)]. Likewise for the force vectors across the central plaquette, no polarization generates any uniformly stable points, but the strength and alignment of the effective forces varies greatly depending on the polarization. 

For the rotationally-symmetric force fields, the self-energy term reaches a minimum at the center of the plaquettes. For an atomic array with sub-wavelength atomic spacing, the oscillatory component of the coherent and dissipative interactions described by~\eref{eqn:couplings} is very small, and so interaction strengths monotonically decrease as one moves away from any single lattice point. As a result, the field minima occur at regions with maximal distance from the lattice points. For a square lattice, these regions are at the center of the plaquette.

The structure of the force fields and the collective energy shifts is more complicated for linearly-polarized arrays, but field minima still appear at the middle of the plaquette. In-plane polarizations result in thin regions of minimal self-energy being formed along lines that point out of each plaquette. We expect the impurity to align its trajectories with these channels, such that it exits the plaquette preferentially in directions aligned with the atom array polarization.

\subsection{Stability analysis based on full dynamics}

\begin{figure}
\centering
\includegraphics[width=0.96\columnwidth]{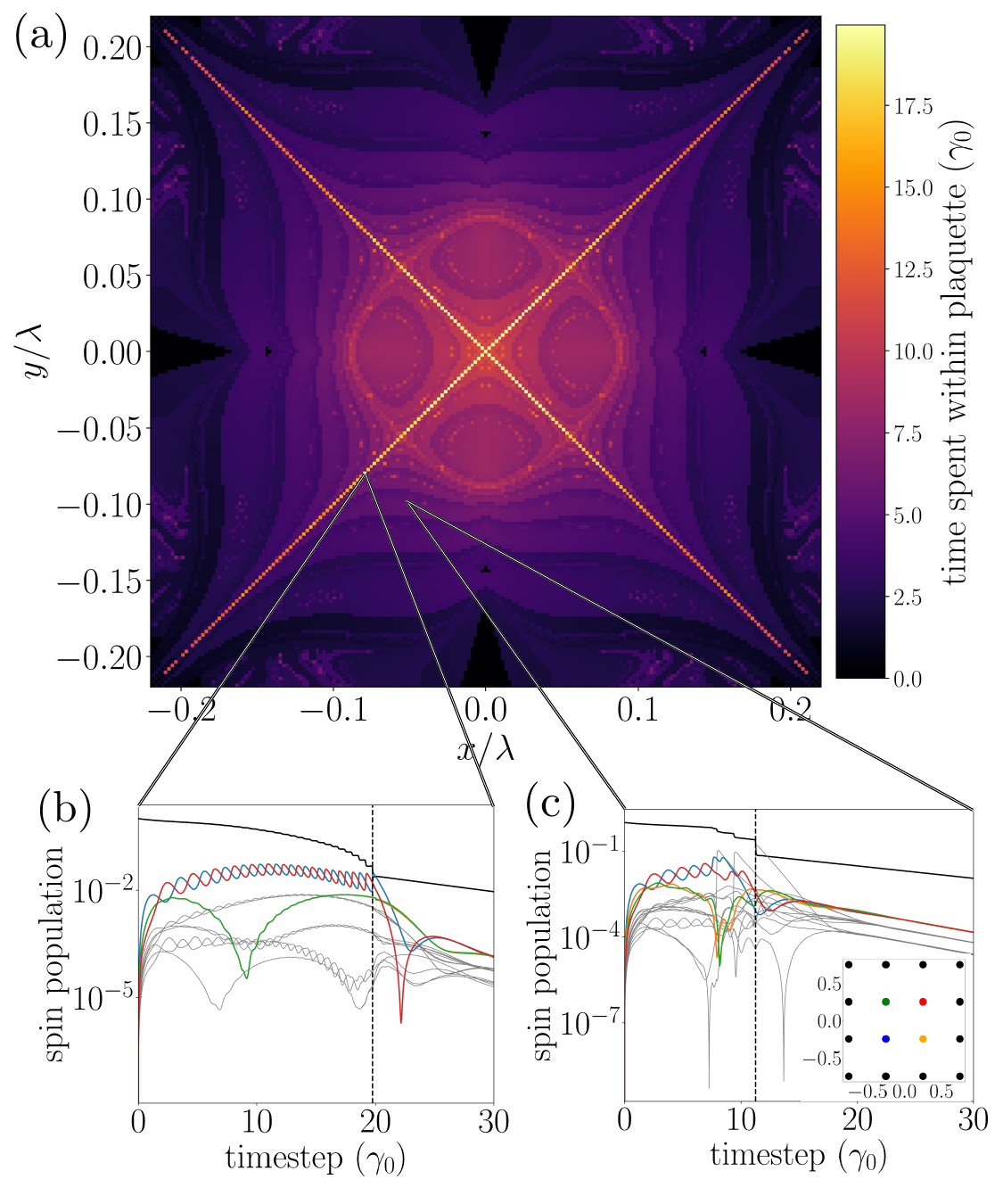}
\caption{
Time the impurity spends in the central lattice plaquette of a $4\times 4$ lattice if it is initialized at all available starting positions within the central plaquette without any initial momentum. We assume circular polarization for both the lattice and impurity atoms. We simulate the impurity's motion within the lattice, and determine how long the impurity remains within the lattice. The time when the impurity exits the lattice is found by identifying the moment when the impurity's spin population decays the most rapidly (see bottom panels). Part (a) of this figure depicts how long the impurity remains within the central plaquette, with the color proportional to the time in units of $\gamma_0$. The detailed spin population curves of the trajectories for two particular initial positions are provided below. A dotted line marks the point of greatest population decay over the trajectory, indicating the time at which the impurity exits the plaquette. Plot (b) on the bottom-left depicts the long-lasting spin population oscillations for an impurity trajectory along the diagonal of the plaquette. Plot (c) on the bottom-right depicts the spin population for an impurity starting off the diagonal. We see that impurities initialized at the off-diagonals exit the lattice much earlier, and also exhibit much shorter oscillatory spin-population dynamics, as this impurity trajectory destabilizes more rapidly.
}
\label{fig:exit_times}
\end{figure}

\begin{figure*}
\centering
\includegraphics[width=\textwidth]{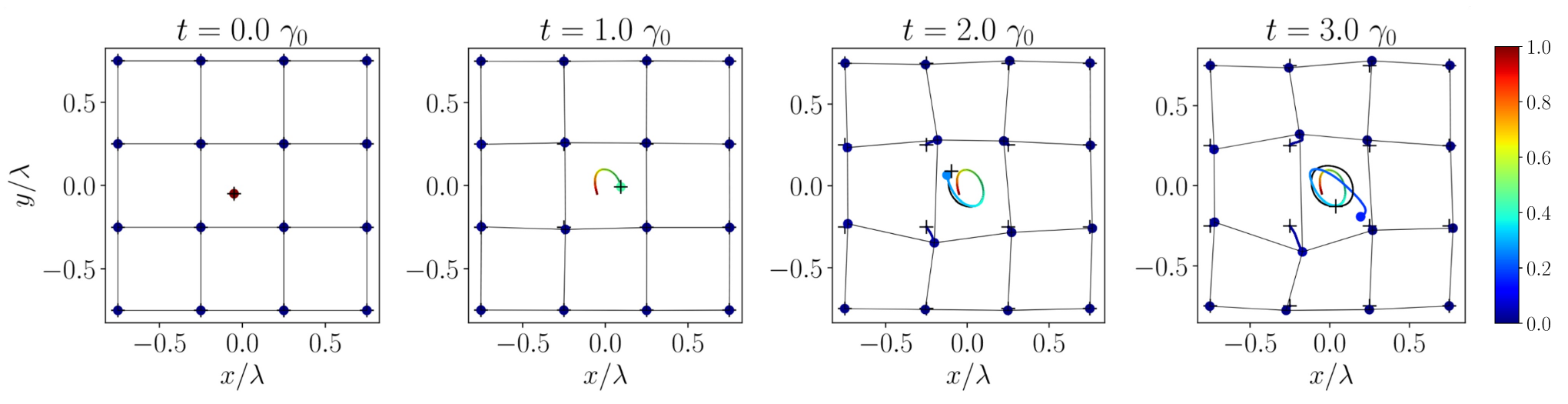}
\caption{Including motional degrees of freedom for the lattice atoms. We add a harmonic trapping potential for the lattice atoms and allow them to move freely within this potential in the semi-classical limit [see~\eref{eqn:lattice_motion}]. Two trajectories are plotted on the lattice. The color coded trajectory indicates the motion of the impurity in the lattice with a weak trapping potential, in which the lattice points can easily move. As in previous figures, the color coding corresponds to the excited state population of the impurity. The other trajectory, denoted by a curved black line, indicates the motion one would obtain for an impurity moving in a lattice where the lattice atoms are pinned down. This trajectory is identical to the trajectory plotted in~\fref{fig:single_traj_example}. The position of the impurity and the lattice points for the fixed lattice is plotted at each individual time by a small cross, while the positions of the lattice points and impurity for the mobile lattice are plotted by small multi-colored circles. The impurity is initialized with an initial momentum $\mathbf{p}_0^I = 0.05(\cos(\theta), \sin(\theta),0)\hbar k_0$ at a position $r^I_0 = (-0.1,-0.1,0)^T a$. The trapping frequency is chosen as $\omega_\mathrm{trap} = 1.0\omega_\mathrm{rec}$. All other parameters are the same as in~\fref{fig:single_traj_example}. While the impurity moves through the lattice, light induced forces between the impurity and the lattice atoms result in a deformation of the underlying lattice structure. This ultimately also influences the impurity dynamics, which can be seen by comparing the solid black line and the color coded line, where the former is the trajectory one would observe in the case where the lattice atoms are pinned to their respective positions. The discrepancy between these curves increases over time once the lattice is deformed significantly.
}  
\label{fig:lattice_motion}
\end{figure*}

Based on these insights from the eliminated lattice potential, we can test how they translate to the full dynamics governed by the set of equations~\eqref{eqn:s_dynamics}-\eqref{eqn:x_p_dyn}. In particular, we are interested in which lattice polarizations, and which regions of the plaquette are associated with the most stable impurity trajectories. In this case we quantify stability as the time the impurity spends within the central lattice region. We solve for the full dynamics by initializing an impurity in its excited state at 625 different initial positions within the central plaquette (a $25 \times 25$ square covering the central plaquette of the lattice), and no initial velocity. Since some of these positions will correspond to unstable configurations within the plaquette the impurity will eventually gain momentum and start to move through the lattice. We time evolve the system until a final time $t_\mathrm{final} = 3.0 \gamma_0$. Each red dot in~\fref{fig:full_stability} corresponds to the recorded impurity position at regular intervals along the atom's trajectory. This implies that regions of high dynamic stability correspond to the dark red regions in~\fref{fig:full_stability}. 

We find that the lines of greatest symmetry within the plaquette attract the impurity the most often. Notably, all distributions of the impurity positions include strong components along the diagonals of the plaquette, because this is where the most robustly stable orbits lie. Along these diagonal axes, the forces imposed by the lattice atoms cancel, and the impurity will not be able to precess in an orbit around the center as illustrated in~\fref{fig:trajectory_example}. Instead, it can just pass through the center of the plaquette, remaining always along a straight trajectory along the diagonal axis.~\fref{fig:full_stability} also nicely illustrates the effect of different dipole moments $\mathbf{d}$. It can be seen that the choice of dipole moment directly influences the structure of the underlying dynamic potential and the regions of high stability change by changing the dipole orientation. The apparent regions of high stability strongly resemble the features of the force fields and energy shifts shown in~\fref{fig:effective_potential}. 

Importantly, for the spherically-polarized case in~\fref{fig:full_stability}(a) the impurity spends more time at the center of the plaquette than the out-of-plane polarized potential~\fref{fig:full_stability}(b), despite them having the same spatial symmetries along the $x$-$y$ plane. A greater density of impurity positions spent outside of the plaquette in~\fref{fig:full_stability}(b) compared to~\fref{fig:full_stability}(a) indicates that, on average, the impurity escapes the plaquette at an earlier time in the out-of-plane polarized array compared to the spherically polarized array. This is likely because the interaction strength of the spherically-polarized dipoles is greater than the interaction strength for out-of-plane polarized dipoles, when averaged over the plaquette plane. Furthermore, the diagonally polarized potential in~\fref{fig:full_stability}(d) keeps the impurity at the palquette center longer than the $x$-polarized potential in~\fref{fig:full_stability}(c), since the former channels the impurity's motion along the diagonal, so that it cannot escape the plaquette without first running into the lattice points in the corners. 

To analyze the stability of the stable points along the diagonal determined in~\fref{fig:full_stability}, we initialize a sequence of impurity trajectories across a $100 \times 100$ grid of initial stationary positions, covering the central plaquette and determine the time at which the impurity exits the plaquette (see~\fref{fig:exit_times}). This time can be readily obtained by analyzing the excited state population dynamics of the impurity, which exhibits a shard decline once the central lattice region is left (see spin population dynamics in~\fref{fig:exit_times}). In comparing the trajectories that start along the diagonal of the central plaquette to those that begin off this diagonal, the on-diagonal trajectories remain within the lattice for almost an order-of-magnitude longer than the off-diagonal trajectories. In other terms, the exit time of the impurity depends very sensitively on the impurity's motion beginning exactly along the diagonal, and then remaining there for as long as possible. This demonstrates a crucial lesson for dynamically trapping impurities within an atomic lattice: if the impurity cannot be already positioned at exactly the center of the plaquette, then it is essential that it at least be positioned along a diagonal of the plaquette, where its oscillatory motion can be dampened until it is trapped at the center. Any off-diagonal motion will lead to swift spin population decay, until the impurity's unstable trajectory leads it rapidly out of the lattice. This explains why the distribution of impurity positions in \fref{fig:full_stability} is so much denser along the diagonal, at least for lattice polarizations that do not pick out a preferred axis. 

We also see in \fref{fig:exit_times} a distinct circular region at the center in which the lattice retains the impurity for a notably longer time than anywhere other than the diagonals. This presumably corresponds to the region where the impurity can move slowly enough through the lattice's potential for it to retain an almost circular orbit within the central plaquette.

\section{Including lattice motion}
So far we have only considered motional degrees of freedom for the impurity, while the lattice atoms were assumed to be pinned to their respective positions. The latter is of course an assumption that does not hold  in realistic experimental setups where a finite trap depth for the lattice atoms will always result in some motion. In particular, the trapping frequency for the lattice atoms can in general also be tuned such that the lattice atoms acquire motional degrees of freedom as well. To include motion  of the lattice atoms within their respective trapping potentials, we add the term, 
\begin{equation}
H_\mathrm{motion}^L = H_\mathrm{kin}^L + H_\mathrm{pot}^L = \sum_{i=1}^{N_L}\left[\frac{(\mathbf{p}^L_i)^2}{2m_L} + \frac{m_L \omega_\mathrm{trap}^2(\mathbf{r}_i^L)^2}{2}\right]
\end{equation}
to the original Hamiltonian~\eqref{eqn:Hamiltonian}, where $m_L$ denotes the mass of the lattice atoms and $\omega_\mathrm{trap}$ is the trapping frequency, which we assume to be equal for all lattice atoms. Introducing these Hamiltonian terms results in an additional set of Heisenberg equations of motion for the lattice atoms' positions $\mathrm{r}_i^L$ and momenta $\mathbf{p}^L_i$, that take the form
\begin{align}
\partial_t{\hat{p}}_{i,\tau}^L = \sum_{j=1,j\neq i}^{N_L} \Big\{ \tilde{\sigma}_i^\dagger \left[\partial_\tau \mathcal{C}_{ij}(\mathbf{r}^L_i,\mathbf{r}_j) \right] \tilde{\sigma}_j + \mathrm{h.c.}\Big\} \nonumber\\ 
+\ \tilde{\sigma}_i^\dagger \left[\partial_\tau \mathcal{C}_{iI}(\mathbf{r}_i,\mathbf{r}^I) \right] \tilde{s} + \mathrm{h.c.} \nonumber \\ 
+ m_L \omega_\mathrm{trap}^2 \mathbf{r}_i^L,
\label{eqn:lattice_motion}
\end{align}
along with $\partial_t \hat{\mathbf{r}}^L_i = \hat{\mathbf{p}}^L_i/m$. 
We show an example of the resultant dynamics in~\fref{fig:lattice_motion}. We initialize an excited impurity at a lattice location within the central plaquette of the lattice, with an initial momentum $\mathbf{p}^I_0 = 0.05 [\cos(\theta),\sin(\theta),0]^T\hbar k_0$ with $\theta = 0.58\pi$, just as in~\fref{fig:single_traj_example}. While the impurity starts to move through the lattice plane on a similar trajectory as in~\fref{fig:single_traj_example}, the lattice structure starts to deform over time due to the forces imposed by the photon exchange between the impurity and the lattice atoms and between the different lattice atoms respectively.

Importantly, the lattice dynamics also influence the dynamics of the impurity. This can be seen by comparing the color coded trajectory in~\fref{fig:lattice_motion} with the thin black line plotted on top of it. The latter shows the dynamics one would get for the same initialization if the lattice atoms were pinned to their initial locations. The deviation of the trajectories increases over time as the lattice motion exerts greater and greater influence on the impurity's dynamics. Hence, the impurity motion gets ``dressed'' by the lattice dynamics. This strongly resembles the original idea of a polaron as proposed by Landau for a charged electron moving through an atomic lattice in solid state structures. This could render the setup studied and presented here a promising platform to study and reproduce complex collective dynamical phenomena known from condensed matter physics in artificial dynamic potentials generated by light.

\section{Conclusions and Outlook}\label{sec:conclusion}
We have investigated the dynamics of an impurity moving in a two dimensional square lattice with subwavelength lattice spacing, and showed that it is a good paradigmatic system to study and analyze the intricate dynamics observed in extended cooperative emitter arrays with light-induced dipole-dipole coupling. The spin population transfer between the impurity and the lattice atoms results in a countervailing force that drives the impurity towards the center of the lattice.

By eliminating the lattice degrees of freedom we illustrated that the experienced force resembles a self-energy for the uncharged impurity. The emergence of such a force could potentially be tied to the emergence of a dynamic artificial gauge field during the dynamics~\cite{dalibard_colloquium_2011, kiffner_abelian_2013, cesa_artificial_2013}. A detailed analysis of this relationship is beyond the scope of the present work, but it renders an interesting avenue for future research.

Another interesting avenue suggested by the findings presented here is whether adding a coherent drive field could result in efficient ground state cooling schemes for the impurity. This is similar to the ideas outlined for cooling lattice vibrations in~\cite{palmer_enhancing_2010, rubies-bigorda_collectively_2024}. Studying such a setup would require going beyond the semiclassical approximation used in our model, which renders a promising avenue for future research.

\acknowledgements
The numerical results were obtained using the \textsc{Quantumoptics.jl} package~\cite{kramer_quantumopticsjl_2018}. 
We acknowledge NSF via PHY-2207972 and via the CUA PFC PHY-2317134.

\appendix

\newpage
\begin{widetext}
\section{Properties of the Green's tensor for a point dipole in free space}\label{app:Greens_Tensor}

The Green's tensor of a point dipole in the electromagnetic vacuum of free space is given in Cartesian coordinates as
\begin{equation}
G_{\alpha\beta}(\mathbf{r}, \omega) = \frac{e^{i\omega r}}{4\pi r} \left[ \left( 1 + \frac{i}{\omega r} - \frac{1}{\omega^2 r^2} \right) \delta_{\alpha\beta} - \left( 1 + \frac{3i}{\omega r} - \frac{3}{\omega^2 r^2} \right) \frac{\mathbf{r}_\alpha \mathbf{r}_\beta }{r^2} \right] - \frac{\delta(r)}{3\omega^2} \delta_{\alpha\beta},
\label{eqn:Green}
\end{equation}
where $\omega = k c$, $r = |\mathbf{r}|$ and $\alpha,\beta = x,y,z$. This Green's tensor determines the interaction strength between pairs of two-level systems via the relation given in~\eref{eqn:couplings}.

The equations of motion for the impurity momentum $\hat{\mathbf{p}}^I$ contain the spatial derivatives of the complex coupling constants $\mathcal{C}_{Ii}$, which corresponds to the spatial derivatives of the Green's tensor given in~\eref{eqn:Green}. The derivative along the $x$-axis is given as
\begin{multline}   
    \frac{\partial}{\partial x} G_{\alpha\beta}(r,\omega) = \frac{\delta_{\alpha\beta}}{4\pi} e^{i \omega r} \left( -\frac{x(-1+i\omega r)}{r^3} - \frac{x(2i + \omega r)}{r^4\omega} + \frac{x(3-i\omega r)}{r^5 \omega^2} \right) \\ 
    - \frac{\delta_{x\alpha\beta}}{4\pi} e^{i\omega r} \left( \frac{x(2r^2 - 3x^2 + i \omega r x^2)}{r^5} + \frac{3 i x (2r^2 - 4x^2 + i \omega r x^2)}{r^6 \omega} - \frac{3x(2r^5 - 5x^2 + i \omega r x^2)}{r^7 \omega^2} \right) \\ 
    - \frac{(1-\delta_{x\alpha\beta})(\delta_{x\alpha})}{4\pi} e^{i\omega r} \left( \frac{\mathbf{r}_\beta (r^2 - 3x^2 + i \omega r x^2)}{r^5} + \frac{3 i \mathbf{r}_\beta (r^2 - 4x^2 + i \omega r x^2)}{r^6 \omega} - \frac{3 \mathbf{r}_\beta (r^2 - 5x^2 + i \omega r x^2)}{r^7 \omega^2} \right) \\ 
    - \frac{(1-\delta_{x\alpha\beta})(\delta_{x\beta})}{4\pi} e^{i\omega r} \left( \frac{\mathbf{r}_\alpha (r^2 - 3x^2 + i \omega r x^2)}{r^5} + \frac{3 i \mathbf{r}_\alpha (r^2 - 4x^2 + i \omega r x^2)}{r^6 \omega} - \frac{3 \mathbf{r}_\alpha (r^2 - 5x^2 + i \omega r x^2)}{r^7 \omega^2} \right) \\ 
    - \frac{(1-\delta_{x\alpha})(1-\delta_{x\beta})}{4\pi} e^{i\omega r} \left(\frac{\mathbf{r}_\alpha \mathbf{r}_\beta x (-3 + i\omega r) }{r^5} - \frac{3\mathbf{r}_\alpha \mathbf{r}_\beta x(4i+r\omega)}{r^6 \omega}+ \frac{3\mathbf{r}_\alpha \mathbf{r}_\beta x (5 - ir\omega)}{r^7 \omega^2}\right),
    \label{eqn:derivative_x}
\end{multline}
where again $\alpha,\beta = x,y,x$. The derivative along the $y$-axis can be obtained via the replacement $x\rightarrow y$ in~\eref{eqn:derivative_x}. The spatial dependence of these derivatives determines the highly non-trivial force fields experienced by the impurity while moving through the array.

\section{Numerical solution for the impurity dynamics}
\label{app:num_solution}
To numerically solve for the coupled dynamics of the spin- and motional degrees of freedom we time evolve the spin wavefunction via the Schrödinger equation $i\partial_t \ket{\psi(t)} = H_\mathrm{spin} \ket{\psi(t)}$ for the non-Hermitian Hamiltonian $H_\mathrm{spin} = H_L + H_I + H_\mathrm{int}^L + H_\mathrm{int}^{LI}$. At each time step this wavefunction is then used to calculate the expectation values $\langle \tilde{s}^\dagger \tilde{\sigma}_i\rangle$ and $\langle \tilde{\sigma}_i^\dagger \tilde{s}\rangle$ in the equations of motion for the impurity~\eqref{eqn:x_p_dyn}. The updated positions after a time step $\Delta t$ obtained from this set of equations are then used to update the coherent and dissipative interactions $J_{ij}$ and $\Gamma_{ij}$ in $H_\mathrm{spin}$ respectively for the next time step. This iterative dynamic procedure effectively solves the full semi-classical dynamics illustrated by the set of equations~\eqref{eqn:sigma_dyn}-\eqref{eqn:x_p_dyn}.

\end{widetext}

\bibliography{references}

\end{document}